\documentclass[journal,onecolumn,10pt,draftclsnofoot]{IEEEtran}
\usepackage{comment}
\usepackage[utf8]{inputenc}
\usepackage{tabularx}
\usepackage{hyperref}
\usepackage{stackrel}
\usepackage{amssymb}
\usepackage{amsmath}
\usepackage{mathtools}
\usepackage[dvips]{graphics}
\usepackage{graphicx}
\usepackage{psfrag}
\usepackage{epsf}
\usepackage{dsfont}
\usepackage{stackrel}
\usepackage{theorem}
\usepackage{bigints}
\usepackage{arydshln}
\usepackage{fancyvrb}
\usepackage{geometry}
\usepackage{pstool}
\usepackage{caption}
\usepackage{subcaption}
\newtheorem{remark}{Remark}

\newtheorem{definition}{Definition}
\newtheorem{theorem}{Theorem}
\newtheorem{proposition}{Proposition}
\newtheorem{lemma}{Lemma}

\newcommand{\defeq}{ \triangleq }

\newcommand{\Markov}{\leftrightarrow}

\begin{document}
\title{Extracting analytic proofs from numerically solved Shannon-type Inequalities}
\author{Ido B. Gattegno and Haim H. Permuter}
\date{\today}
\maketitle
\vspace*{-15mm}
\section{Introduction}
A class of information inequalities, called Shannon-type inequalities (STIs), can be proven via a computer software called ITIP \cite{yeung_itip}. In previous work \cite{gattegno2016fourier}, we have shown how this technique can be utilized to Fourier-Motzkin elimination algorithm for Information Theoretic Inequalities. Here, we provide an algorithm for extracting \emph{analytic} proofs of information inequalities. Shannon-type inequalities are proven by solving an optimization problem. We will show how to extract a \emph{formal} proof of numerically solved information inequality. Such proof may become useful when an inequality is implied by several constraints due to the PMF, and the proof is not apparent easily. More complicated are cases where an inequality holds due to both constraints from the PMF and due to other constraints that arise from the statistical model. Such cases include information theoretic capacity regions, rate-distortion functions and lossless compression rates. We begin with formal definition of Shannon-type information inequalities. We then review the optimal solution of the optimization problem and how to extract a proof that is readable to the user.
\section{Preliminaries and Notations}
\par We use the following notation.
Calligraphic letters denote discrete sets, e.g., $\mathcal{X}$. The empty set is denoted by $\phi$, while $\mathcal{N}_n\defeq\{1,2,\dots,n\}$ is a set of indices.
Lowercase letters, e.g. $x$, represent variables. A vector of $n$ variables $(x_1,\dots,x_n)$ is denoted by $\mathbf{x}_{\mathcal{N}_n}$, and its substring as $\mathbf{x}_\alpha=(x_i\in\Omega\;|\;i\in\alpha,\;\phi\neq\alpha\subseteq\mathcal{N}_n)$, e.g., $\mathbf{x}_{\{1,2\}}=(x_1,x_2)^\top$;
whenever the dimensions are clear from the context, the subscript is omitted.
Vector inequalities, e.g., $\mathbf{v}\geq\mathbf{0}$, are in element-wise sense.
Random variables are denoted by uppercase letters, e.g., $X$, with similar conventions for random vectors.

\section{Information inequalities and constraints}\label{itip}
In \cite{yeung2008information}, Yeung characterized a subset of information inequalities named \textit{Shannon-type inequaliteis} (STIs), that are provable using a computer program called ITIP \cite{yeung_itip}. More work on the ITIP was done in \cite{xitip}. This section consist of the mathematical review of this work. We establish a \textit{canonical form} for linear combination of Shannon's information measures, which uniquely represent the expression as a linear combination of joint entropies. By giving constraints of non-negativity on the information measures, we establish a region where linear information inequalities residue. A theorem provides a minimization problem, which can be solved by linear programming techniques, makes the identification of true information inequalities applicable.
\subsection{Unconstrained inequalities}\label{itip_unconst}
\par Given a random vector $\mathbf{X}_{\mathcal{N}_n}$ that take values in $\mathcal{X}_1\times\cdots\times\mathcal{X}_n$,
define $\mathbf{h}_\ell\defeq\big(H(\mathbf{X}_\alpha\big)|\phi\neq\alpha\subseteq\mathcal{N}_n)$\footnote{We assume a lexicographical ordering of the elements of $\mathbf{h}_\ell$.}.
Let $\mathcal{P}$ be the set of all probability mass functions (PMFs) over $\mathcal{X}_1\times\cdots\times\mathcal{X}_n$.
Moreover, for every $p\in\mathcal{P},\;\mathbf{h}_l(p)\in\mathbb{R}^{2^n-1}$ is a vector whose entries are the values of $H(\mathbf{X}_\alpha), \phi\neq\alpha\subseteq\mathcal{N}_n$,
with respect to $p$.
\begin{definition}[Basic information measure (BIM)]\label{itip_unconst_def_basic}
	An information measure is called \textit{basic} if it takes on one of the following forms:
	\begin{subequations}\label{itip_unconst_eq_basic}
		\begin{align}
			&H(\mathbf{X}_\alpha|\mathbf{X}_\gamma)\\
			&I(\mathbf{X}_{\alpha};\mathbf{X}_{\beta}|\mathbf{X}_\gamma),
		\end{align}
	\end{subequations}
	where $\alpha,\beta,\gamma\subseteq\mathcal{N}_n$ and $\alpha,\beta\neq\phi$.
\end{definition}
\begin{definition}[Elemental information measure (EIM)]\label{itip_unconst_def_elemental}
	An information measure is called \textit{elemental} if it takes on one of the following forms:
	\begin{subequations}\label{itip_unconst_eq_elemental}
		\begin{align}
			&H(X_i|\mathbf{X}_{\mathcal{N}_n\backslash\{ i\}})\\
			&I(X_i;X_j|\mathbf{X}_{\mathcal{K}}),
		\end{align}
		where $i,j\in\mathcal{N}_n,\;i\neq j,\; \mathcal{K}\subseteq\mathcal{N}_n\backslash \{i,j\}$
	\end{subequations}
\end{definition}
\begin{lemma}\label{itip_unconst_lemma_basic_to_elemental}
	Every BIM can be represented as a \textit{linear combination} of EIMs with non-negative coefficients.
\end{lemma}
\par By the definition of mutual information and by the entropy chain rule, for every $i,j\in\mathcal{N}_n,\;i\neq j$ and $\mathcal{K}\subseteq\mathcal{N}_n\backslash \{i,j\}$, we have
\begin{subequations}\label{itip_unconst_eq_chainrules}
	\begin{align}
		H(X_i|\mathbf{X}_{\mathcal{N}_n\backslash \{i\}})=&H(\mathbf{X}_{\mathcal{N}_n})-H(\mathbf{X}_{\mathcal{N}_n\backslash \{i\}}), \\
		I(X_i;X_j|\mathbf{X}_{\mathcal{K}})=&H(X_i,\mathbf{X}_{\mathcal{K}})+
		H(X_j,\mathbf{X}_{\mathcal{K}})\\&-H(X_i,X_j,\mathbf{X}_{\mathcal{K}})-H(\mathbf{X}_{\mathcal{K}}). \nonumber
	\end{align}
\end{subequations}
\par Lemma \ref{itip_unconst_lemma_basic_to_elemental} combined with (\ref{itip_unconst_eq_chainrules}) implies that every BIM is uniquely\footnote{For the proof of uniqueness see \cite[Section 13.2]{yeung2008information}.} representable as a linear combination of unconditional joint entropies.
This representation, which is called the \textit{canonical form}, allows one to write every linear combination of BIMs as $\mathbf{b}^\top\mathbf{h}_\ell$, where $\mathbf{b}$ is a vector of coefficients\footnote{Henceforth, an arbitrary linear combination of BIMs is denoted by $\mathbf{b}^\top\mathbf{h}_\ell$.}.
\begin{definition}
	An information inequality $\mathbf{b}^\top\mathbf{h}_\ell\geq0$
	\textit{always holds} if $\mathbf{b}^\top\mathbf{h}_\ell(p)\geq0$, for every $p\in\mathcal{P}$.
\end{definition}
\begin{proposition}\label{itip_unconst_corollary_opt_in_gamma_star}
	An information inequality $\mathbf{b}^\top\mathbf{h}_\ell\geq0$ always holds if and only if (iff)
	\begin{subequations}\label{itip_unconst_eq_opt_in_gamma_star}
		\begin{align}
			\min_{p\in\mathcal{P}}\mathbf{b}^\top\mathbf{h}_\ell(p)=\min_{\mathbf{h}_\ell(p)\in\Gamma_n^*}\mathbf{b}^\top\mathbf{h}_\ell(p)=0,
		\end{align}
		where
		\begin{align}
			\Gamma_n^*=\bigcup_{p\in\mathcal{P}}\mathbf{h}_\ell(p).
		\end{align}
	\end{subequations}
\end{proposition}
Proposition \ref{itip_unconst_corollary_opt_in_gamma_star} follows since there is always a $p\in\mathcal{P}$ for which $\mathbf{b}^\top\mathbf{h}_\ell(p)=0$.
\par The optimization problem in (\ref{itip_unconst_eq_opt_in_gamma_star}) is infeasible as it involves optimizing over the set of all PMFs of $n$ discrete random variables. Therefore, an algorithm that numerically proves information inequalities requires an simpler alternative description of $\Gamma_n^*$. Such a description, being currently unknown, leads one to search for a different subspace of $\mathbb{R}^{2^n-1}$, that is, in a sense, similar to $\Gamma_n^*$, based on which numerical proofs can be implemented.
\begin{definition}[Basic and elemental inequalities]\label{itip_unconst_def_basic_elem_inequalities}
	Non-negativity inequalities on BIMs and EIMs are called \textit{basic inequalities} (BIs) and \textit{elemental inequalities} (EIs), respectively.
\end{definition}
\par Every $\mathbf{h}\in\Gamma_n^*$ is a vector of entropies that is induced by some $p\in\mathcal{P}$, and, in particular, satisfies all BIs.
Since BIs are linear constraints on $\mathbf{h}$, in \cite{yeung2008information},
\begin{align}\label{itip_unconst_eq_gamman_words}
	\Gamma_n\defeq\{\mathbf{h}\in\mathbb{R}^{2^n-1}|\;\mathbf{h}\;\text{satisfies all BIs}\}
\end{align}
was proposed as an alternative for $\Gamma_n^*$.
\begin{lemma}[Minimality of elemental inequalities]\label{itip_unconst_lemma_minimality_of_elem}
	The set of EIs is minimal in sense that every BI is implied by a subset of EIs.
\end{lemma}
\begin{remark}\label{itip_unconst_remark_amountIMs}
	There are $n+\binom{n}{2}2^{n-2}$ EIs while the amount of BIs is bounded by $\sum_{j=1}^n\binom{n}{j}\Big(2^{n-j}+\sum_{i=1}^{n-j}\binom{n-j}{i}2^{n-j-i}\Big)$ from below.
\end{remark}
Based on Remark \ref{itip_unconst_remark_amountIMs} and Lemma \ref{itip_unconst_lemma_minimality_of_elem}, we write
\begin{align}\label{itip_unconst_eq_gamma_n_by_G}
	\Gamma_n=\{\mathbf{h}\in\mathbb{R}^{2^n-1}\;|\;\rm{G}\mathbf{h}\geq\mathbf{0}\},
\end{align}
where G is a matrix such that the elements of $\rm{G}\mathbf{h}_\ell$ are all EIMs.
\begin{theorem}\label{itip_unconst_theorem_always_holds_gamma_n}
	Let $\mathbf{b}^\top\mathbf{h}_\ell\geq0$ be an information inequality, and let
	\begin{align}\label{itip_unconst_eq_always_holds_gamma_n}
		\rho^*=\min_{\substack{\mathbf{h}:\\ \mathrm{G}\mathbf{h}\geq\mathbf{0}}}\mathbf{b}^\top\mathbf{h}.
	\end{align}
	If $\rho^*=0$ then $\mathbf{b}^\top\mathbf{h}_\ell\geq0$ always holds.
\end{theorem}
\par
The proof of Theorem \ref{itip_unconst_theorem_always_holds_gamma_n} follows from Proposition \ref{itip_unconst_corollary_opt_in_gamma_star} since every $\mathbf{h}\in\Gamma_n^*$ satisfies all EIs, which implies that $\Gamma_n^*\subseteq\Gamma_n$\footnote{$\Gamma_2^*=\Gamma_2$ but $\Gamma_3^*\neq\Gamma_3$\cite[Section 15.1]{yeung2008information}.}.
The optimization problem (\ref{itip_unconst_eq_always_holds_gamma_n}) is solvable using \textit{linear programming} (LP) optimization methods \cite{schrijver1998theory}.
Information inequalities that are provable by Theorem \ref{itip_unconst_theorem_always_holds_gamma_n} form a subset of inequalities called \textit{unconstrained Shannon-type inequalities (STIs)}.
\subsection{Constrained STIs}\label{itip_const}
Some information inequalities (respectively, identities) hold only when a certain structure is imposed on the PMFs.
Such a structure may account for independencies between random variables, Markov chains and functional dependencies. We formulate these constraints on the PMF domain as linear constraints on entropy and mutual information terms.

\begin{lemma}\label{itip_const_lemma_const_to_entropies_consts}
	Let $\{X_i\}_{i=1}^n$ be a collection of random variables.
	\begin{enumerate}
		\item $\{X_i\}_{i=1}^n$ are mutually independent iff $H(\mathbf{X}_{\mathcal{N}_n})=\sum_{i=1}^nH(X_i)$.
		\item $\{X_i\}_{i=1}^n$ are pairwise independent iff $I(X_i;X_j)=0$, for every $1\leq i\neq j\leq n$.
		\item Let $\alpha,\beta\subseteq\mathcal{N}_n$ and $\alpha,\beta\neq\phi$. $\mathbf{X}_\alpha$ is a function of $\mathbf{X}_\beta$ iff $H(\mathbf{X}_\alpha|\mathbf{X}_\beta)=0$.
		\item Let $\alpha,\beta,\gamma,\delta\subseteq\mathcal{N}_n$ and $\alpha,\beta,\gamma,\delta\neq\phi$. $\mathbf{X}_\alpha-\mathbf{X}_\beta-\mathbf{X}_\gamma-\mathbf{X}_\delta$ forms a Markov chain iff $I(\mathbf{X}_\alpha;\mathbf{X}_\gamma,\mathbf{X}_\delta|\mathbf{X}_\beta)=0$ and $I(\mathbf{X}_\alpha,\mathbf{X}_\beta;\mathbf{X}_\delta|\mathbf{X}_\gamma)=0$\footnote{Property 4 of Lemma \ref{itip_const_lemma_const_to_entropies_consts} is an extension of \cite[Section 13.3.2]{yeung2008information} and can be generalized to account for longer Markov chains.}.
	\end{enumerate}
\end{lemma}
\par Every set of constraints as in Lemma \ref{itip_const_lemma_const_to_entropies_consts} is representable by 
\begin{align}
	\mathrm{Q}\mathbf{h}_\ell=\mathbf{0}
\end{align}
where $\mathrm{Q}$ is a matrix whose rows are the coefficients that correspond to each constraint.
\begin{proposition}\label{itip_const_prop_pmf}
	Given a PMF defined on discrete random variables, the constraints which induced by reading the joint PMF defines all probabilistic relations between the random variables.
\end{proposition}
\par For instance, given a PMF $P(X,Y,Z)$ where $X-Y-Z$ forms a Markov chain, we can write
\begin{align}
	P(X,Y,Z)=P(X)P(Y|X)\hspace*{-7mm}\underbrace{P(Z|Y)}_{H(Z|Y)=H(Z|Y,X)}.
\end{align}
The induced constraint is equivalent to $I(X;Z|Y)=0$, which implies that $X-Y-Z$ forms a Markov chain.
By Lemma \ref{itip_const_lemma_const_to_entropies_consts}, all such constraints imply the joint PMF.
\begin{theorem}\label{itip_const_theorem_const_STI}
	Let $\mathbf{b}^\top\mathbf{h}_\ell\geq0$ be an information inequality and
	\begin{align}
		\rho^*=\min_{\substack{\mathbf{h}:\\ \rm{G}\mathbf{h}\geq\mathbf{0}\\
				\rm{Q}\mathbf{h}=\mathbf{0}}}\mathbf{b}^\top\mathbf{h}.
	\end{align}
	If $\rho^*=0$ then $\mathbf{b}^\top\mathbf{h}_\ell\geq0$ always holds under the constraints $\rm{Q}\mathbf{h}_\ell=\mathbf{0}$.
\end{theorem}
\par Constrained information inequalities that are captured by Theorem \ref{itip_const_theorem_const_STI} are called \textit{constrained STIs}.
\section{Optimization problems and optimal solution}
	\subsection{The Lagrangian dual function}
		An LP problem is a private case of convex optimization problems. The reader may refer to \cite{boyd2004convex} for further study about convex optimization (here we use lemmas from Chapter 5).
		Consider an optimization problem in the standard form
		\begin{subequations}\label{FIP: Dual, Convex Original problem}
		\begin{align}
		\text{minimize: }	&f_0(\mathbf{x}) \\
		\text{subject to: }	&f_i(\mathbf{x}) \leq 0,\qquad i=1,\dots m \\
							&h_i(\mathbf{x}) = 0,\qquad i=1,\dots,p
		\end{align}
		\end{subequations}
		with $\mathbf{x}\in\mathbb{R}^n$. The function $f_0(\mathbf{x})$ is called the \emph{objective function}. \\
		We define $\mathcal{D}$ to be the domain of this problem,
		\begin{align*}
			\mathcal{D}\defeq \left\{\bigcap_{i=0}^m{dom}(f_i)\right\}\bigcap \left\{\bigcap_{i=1}^p{dom}(h_i)\right\}
		\end{align*}
		where $dom(\cdot)$  is the domain of its arguments. We assume nonempty domain, i.e., $\mathcal{D} \neq \phi$ and that there is an optimal solution for this problem.
	\begin{definition}[Tha Lagrangian]
		The Lagrangian $L:\mathbb{R}^n\times\mathbb{R}^m\times \mathbb{R}^p \to \mathbb{R}$ associated with the problem in (\ref{FIP: Dual, Convex Original problem}) is
		\begin{align}
			L(\mathbf{x},\mathbf{\lambda,\mathbf{\nu}}) \defeq f_0(\mathbf{x}) + \sum_{i=1}^{m}\lambda_if_i(\mathbf{x}) + \sum_{i=1}^{p}\nu_ih_i(\mathbf{x})
		\end{align}
		where $\mathbf{x}\in \mathbb{R}^n,\;\mathbf{\lambda}\in \mathbb{R}^m$ and $\mathbf{\nu}\in \mathbb{R}^p$.
		We refer $\lambda_i$ as the \emph{Lagrange multiplier} of the $i$-th inequality constraint and $\nu_i$ as the \emph{Lagrange multiplier} of the $i$-th equality constraint.
	\end{definition}
	\begin{definition}[Lagrangian dual function]
		The Lagrangian dual function  $g:\mathbb{R}^m\times\mathbb{R}^p\to\mathbb{R}$ is the infimum of the Lagrangian over $\mathbf{x}$:
		\begin{align*}
			g(\mathbf{\lambda},\mathbf{\nu})\defeq \inf_{\mathbf{x}\in\mathcal{D}} L(\mathbf{x},\mathbf{\lambda,\mathbf{\nu}})
		\end{align*}
	\end{definition}
	\begin{lemma}\label{FIP: Lemma, dual function lower that p star}
		For any $\mathbf{\lambda} \geq \mathbf{0}$ and any $\mathbf{\nu}$,
		\begin{align}
			g(\mathbf{\lambda},\mathbf{\nu}) \leq p^\ast
		\end{align}
		where $p^\ast$ is the optimal solution of the problem in (\ref{FIP: Dual, Convex Original problem}).
	\end{lemma}
	Let $d^\ast$ be optimal solution of the following optimization problem
	\begin{subequations}\label{FIP: Dual, Convex dual problem}
		\begin{align*}
		\text{maximize: }&\ \ \ g(\mathbf{\lambda},\mathbf{\nu}) \\
		\text{subject to:}&\ \ \ \mathbf{\lambda}\geq 0.
	\end{align*}
	\end{subequations}
	Note that since the problem in (\ref{FIP: Dual, Convex Original problem}) is convex, (\ref{FIP: Dual, Convex dual problem}) is also convex. We refer $\mathbf{\lambda}^\ast,\mathbf{\nu}^\ast$ as the \emph{optimal Lagrange multipliers} and $b^\ast$ as the \emph{dual optimal solution} of the problem in (\ref{FIP: Dual, Convex dual problem}). In general, by Lemma \ref{FIP: Lemma, dual function lower that p star} we know that $d^\ast \leq p^\ast$. Furthermore, under specific conditions we achieve equality.
	\subsection{Strong duality}
	When $d^\ast = p^\ast$, the solutions of both  the dual and original optimization problems coincide. If that is the case, we say we have \emph{strong duality}. 
	We here assume all equality constraints are affine\footnote{In many information theoretic problems, the inequalities are affine functions of the entropies.}. Thus, the equality constraints in (\ref{FIP: Dual, Convex Original problem}) can be replaced with $\mathrm{A}\mathbf{x} = \mathbf{0}$.
	\begin{definition}[Slater's condition (for affine constraints)]
		Assume that $f_i(\mathbf{x})$ are affine functions of $\mathbf{x}$ for $i=1,\dots,k$. \\
		If there exists $\mathbf{x}\in relint(\mathcal{D})$ , where $relint(\mathcal{D})$ is the relative interior of $\mathcal{D}$, such that
		\begin{subequations}
		\begin{align}
			f_i(\mathbf{x})&\leq 0, \qquad i=1,\dots k \\
			f_i(\mathbf{x})&< 0, \qquad i=k+1,\dots m \\
			\mathrm{A}\mathbf{x}&=\mathbf{0}
		\end{align}
		\end{subequations}
		then we say that Slater's condition holds.
	\end{definition}
	\begin{lemma}\label{FIP: Lemma, Slater imply strong duality}
		If Slater's condition holds, then \emph{strong duality} exists.
	\end{lemma}
	\begin{remark}
		In LP problems, all constraints are affine, and therefore Slater's condition reduce to weak inequality in all constraints. Moreover, if the solution of the problem is feasible, then Slater's condition holds and we have strong duality.
	\end{remark}
		Consider an LP problem of the form
		\begin{subequations}
			\begin{align}
			\text{minimize: }&\mathbf{c}^\top \mathbf{x} \\
			\text{subject to: }&\mathrm{A}\mathbf{x}=\mathbf{b} \\
			&\mathrm{B}\leq \mathbf{d}
		\end{align}
		\end{subequations}
		The Lagrangian of this problem is
		\begin{align}
			L(\mathbf{x},\mathbf{\lambda},\mathbf{\nu}) = \mathbf{c}^\top\mathbf{x} + \mathbf{\lambda}^\top(\mathrm{B}\mathbf{x}-\mathbf{d})+\mathbf{\nu}^\top(\mathrm{A}\mathbf{x}-\mathbf{b})
		\end{align}
		and the dual function is
		\begin{subequations}
			\begin{align}
			g(\mathbf{\lambda},\mathbf{\nu}) &= \inf_{\mathbf{x}}L(\mathbf{x},\mathbf{\lambda},\mathbf{\nu}) \\
			&=(-\mathbf{b}^\top\mathbf{\nu} - \mathbf{d}^\top\mathbf{\lambda}) + \inf_{\mathbf{x}}\{(\mathbf{c} + \mathrm{B}\mathbf{\lambda} + \mathrm{A}^\top\mathbf{\nu})^\top\mathbf{x}\}
			\end{align}
		\end{subequations}
		subject to $\lambda \geq \mathbf{0}$.
	\begin{lemma}[Optimal Lagrange multipliers of an LP problem]\label{FIP: Lemma, optimal Lagrange multipliers}
		If a solution to an LP problem with linear constraints exists, then
		\begin{align}
			\mathrm{A}^\top\nu^\ast + \mathrm{B}^\top\lambda^\ast+\mathbf{c} = 0
		\end{align}
	\end{lemma}
	The proof of Lemma \ref{FIP: Lemma, optimal Lagrange multipliers} follows directly from the definition of the dual function, since
	\begin{align}
		g(\lambda,\nu) = \begin{cases}
		-\mathbf{b}^\top\nu -\mathbf{d}^\top\lambda& \mathrm{A}^\top\nu + \mathrm{B}^\top\lambda+\mathbf{c} = 0 \\
		-\infty&\text{ otherwise}
		\end{cases}
	\end{align}
	Recall that since Slater's condition holds, $g(\lambda^\ast,\nu^\ast)=p^\ast$.
	Consequently, using the \emph{optimal} Lagrange multipliers, we can represent the \emph{linear} objective function by means of \emph{linear combination} of the constraints.
	\section{Extracting formal proof from the optimal solution}
	Recall from Section \ref{itip_const} that non-negativity of linear combination of information measures can be proven by solving an LP problem.
	Assume we want to prove the following inequality in the canonical form
\begin{align}
	\mathbf{f}_L^\top\mathbf{h}_\ell(p) \leq \mathbf{f}_R^\top \mathbf{h}_\ell(p),\qquad p\in  \mathcal{Q}
\end{align}
where $\mathcal{Q}$ is a subspace of $\mathcal{P}$ where the constraints due to the PMF hold.
Define 
\begin{align}
		\mathbf{f}_D \defeq \mathbf{f}_R - \mathbf{f}_L
\end{align}
The corresponding LP problem we solve to check in it is an unconstrained STI is
\begin{subequations}\label{FIP: inequality prover LP}
	\begin{align}
		\text{minimize: }&\mathbf{f}_D^\top \mathbf{h} \\
		\text{subject to: }& -\mathrm{G}\mathbf{h} \leq \mathbf{0} \\
		 &\ \ \ \ \mathrm{Q}\mathbf{h} = \mathbf{0}
	\end{align}
\end{subequations}
where $\mathrm{G}\mathbf{h}_\ell\geq \mathbf{0}$ represent the elemental inequality and $\mathrm{Q}\mathbf{h}_\ell=\mathbf{0}$ the constraints due to the PMF.
Note that both inequality and equality constraints in (\ref{FIP: inequality prover LP}) are affine. If a solution to that problem exists, by Lemma \ref{FIP: Lemma, optimal Lagrange multipliers}, we have
\begin{align}\label{FIP: inequality equal elemental and constraints}
	\mathbf{f}_D = \mathrm{G}^\top \lambda^\ast - \mathrm{Q}^\top \nu^\ast
\end{align}
Thus, we can represent the coefficients of the objective by rows of $\mathrm{G}$ and $\mathrm{Q}$. Define $\mathbf{G}_\ell(p)$ and $\mathrm{Q}_\ell(p)$ to be vectors with labels correspond to $\mathrm{G}$ and $\mathrm{Q}$, respectively. The labels in the components are information measures which are represented by rows of the corresponding matrix. For instance, assume a case where there are only two variables, $(X_1,X_2)$.
By our definitions,
\begin{subequations}
	\begin{align}
	\mathbf{h}_\ell(p)^\top &= \left[H(X_1),\;H(X_2),\;H(X_1,X_2)\right] \\
	\mathbf{G}_\ell^\top&=\left[H(X_1|X_2),\;H(X_2|X_1),\;I(X_1;X_2)\right] \\
	\mathrm{G}&=\left[\begin{array}{ccc}
	0 &-1 &1 \\
	-1& 0 &1 \\
	1 & 1 & -1
	\end{array}\right]
\end{align}
\end{subequations}
Note that $\mathbf{h}_\ell^\top(p)\mathbf{f}_D$ is the canonical form the RHS minus the LHS of the inequality we aim to prove. Similarly, the components of $\mathbf{h}^\top_\ell(p)\mathrm{G}$ and $\mathbf{h}^\top_\ell(p)\mathrm{Q}$ are the canonical forms of $\mathbf{G}_\ell^\top(p)$ and $\mathbf{Q}^\top_\ell(p)$, respectively.
From (\ref{FIP: inequality equal elemental and constraints}) we have
\begin{subequations}
	\begin{align}
	\mathbf{h}^\top_\ell(p)(\mathbf{f}_R - \mathbf{f}_L) &=  \mathbf{h}^\top_\ell(p)\mathrm{G}\lambda^\ast - \mathbf{h}^\top_\ell(p)\mathrm{Q}\nu^\ast\\
	&=\mathbf{G}_\ell^\top(p)\lambda^\ast-\mathbf{Q}^\top_\ell(p)\nu^\ast
\end{align}
\end{subequations}
We then obtain a representation of the difference between R.H.S and L.H.S by elemental inequality and PMF constraints. From this representation, proving the inequality is more apparent. Since $\lambda\geq \mathbf{0}$ and all elemental information measures are nonnegative, it is easy to see that $\mathbf{G}_\ell^\top(p)\lambda^\ast$ is nonnegative.
As for $\mathbf{Q}^\top_\ell(p)\nu^\ast$, it is a sum of information measures that are each equal to zero due Markov chains induced by the PMF. We refer this representation as the \emph{elemental form} of the expression. Equality between the original representation and the \emph{elemental form} can be shows either using information theoretic \emph{identities} or by showing that the \emph{canonical forms} of both are the same.

\section{An illustration}
We provide an illustration of how the algorithm works and what is the extracted proof. 
\begin{figure}
	\includegraphics[scale=0.5]{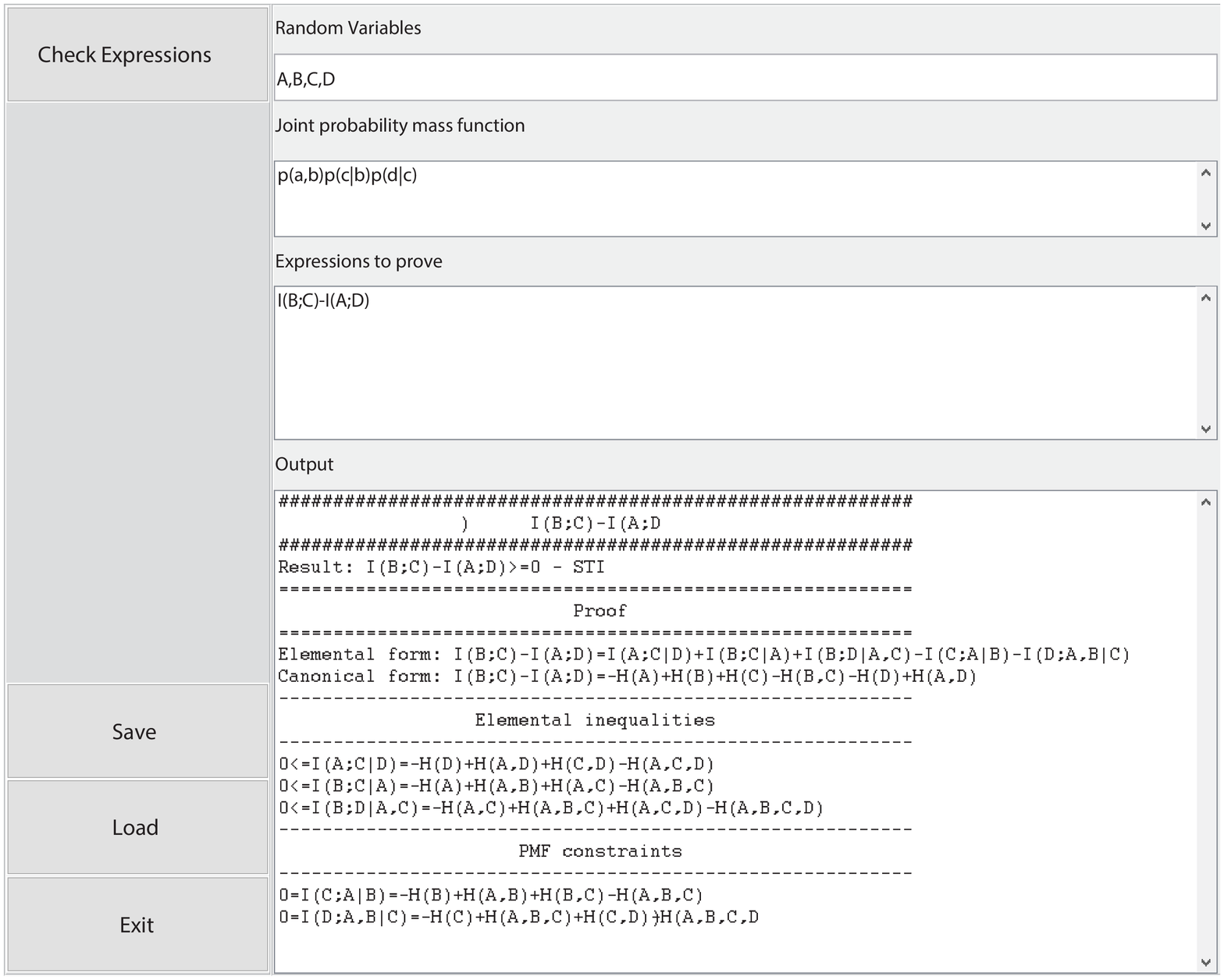}
	\centering
	\caption{Demonstration of the algorithm}
	\label{FIP: Figure, Example}
\end{figure}
Let $(A,B,C,D)$ be $4$ random variables such that $A\Markov B \Markov C \Markov D$ is a Markov chain. The PMF of those variables factorizes as follows
\begin{align}
	P(a,b,c,d)=P(a,b)P(c|b)P(d|c)
\end{align}
Following Proposition \ref{itip_const_prop_pmf}, we obtain the following constraints
\begin{align}
	I(C;A|B)&=0 \\
	I(D;A,B|C)&=0.
\end{align}
Since $A\Markov B \Markov C \Markov D$ is a Markov chain, 
\begin{align}
	I(A;D)\leq I(B;C)
\end{align}
always holds due to the PMF factorization.
Using the canonical form of the R.H.S minus the L.H.S, it can be verified that
\begin{subequations}
	\begin{align}
	I(B;C)-I(A;D) =& I(A;C|D) + I(B;C|A) + I(B;D|A,C)
I(C;A|B) - I(D;A,B|C).
\end{align}
\end{subequations}
This representation clarify  that $I(B;C)-I(A;D) \geq 0$ because of the Markov chains. An implementation\footnote{The algorithm was implemented in MATLAB.} of this algorithm is demonstrated in Fig. \ref{FIP: Figure, Example}. 
\bibliography{IEEEabrv,ref}
\end{document}